\documentclass[aps,prl,reprint,superscriptaddress,longbibliography]{revtex4-2}

\usepackage[pdftex]{graphicx}

\usepackage{dcolumn}
\usepackage{bm}
\usepackage{amsmath}
\usepackage{amssymb}
\usepackage{ulem}
\usepackage{braket}
\usepackage{bbold}

\usepackage[caption=false]{subfig}
\graphicspath{ {img/} }

\usepackage[usenames,dvipsnames]{xcolor}
\usepackage{hyperref}
 \hypersetup{
    colorlinks=true,
    linkcolor=ForestGreen,
    citecolor=RoyalBlue,
    filecolor=BrickRed,
    urlcolor=RoyalBlue,
    bookmarks=true,
    bookmarksopen=true,
    bookmarksnumbered=true
  }

\usepackage{tikz}
\usetikzlibrary{positioning}

\newcommand{\<}{\langle}

\renewcommand{\>}{\rangle}
\renewcommand{\(}{\left(}
\renewcommand{\)}{\right)}
\renewcommand{\[}{\left[}
\renewcommand{\]}{\right]}
\renewcommand{\b}[1]{\mathbf{#1}} 

\newcommand{\bs}[1]{\boldsymbol{#1}}
\renewcommand{\d}{\partial}

\newcommand{\eps}{\epsilon}
\newcommand{\p}{\parallel}

\renewcommand{\Re}{\operatorname{\text{Re}}}

\begin{document}

\title{Velocity collapse and non-conformal spiral phase in the sawtooth spin chain}
\author{Nai Chao Hu}
\affiliation{Department of Physics and Astronomy, Ghent University, Krijgslaan 281, S9, 9000 Gent, Belgium}

\begin{abstract}
Recent matrix-product-state calculations show that the spiral phase in the sawtooth chain has numerical signatures that are difficult to reconcile with an ordinary conformal critical point: a large apparent central charge, slow dynamical scaling, nearly flat excitations, and no detectable dimerization.
We develop a bosonization theory for this phenomenology by embedding the sawtooth limit in a zigzag ladder described by two coupled SU(2)$_1$ conformal field theories characterized by an extreme velocity ratio.
We show that the sawtooth geometry cancels the leading staggered interaction, leaving a marginal twist interaction that selectively collapses the slow apical spin velocity. Crucially, as this velocity vanishes, the generated apical backscattering interaction diverges only in dimensionless units, causing the energy scale to collapse independently of the spatial correlation length. This mechanism naturally accounts for many of the numerical anomalies and we interpret the perturbative flow as an entrance to local quantum criticality in the strong-coupling regime.
\end{abstract}

\maketitle

\paragraph*{Introduction. ---}
The sawtooth, or $\Delta$, spin chain is a paradigmatic frustrated spin-1/2 system where numerical evidence points to low-energy behavior outside the standard conformal-field-theory (CFT) template.
The microscopic model is
\begin{equation}
    H = \sum_j\Big[J_2\,\b{S}_j\!\cdot\!\b{S}_{j+1}
    + J_1\,\bs{\tau}_j\!\cdot\!(\b{S}_j+\b{S}_{j+1})\Big],
    \label{eq:H_sawtooth_draft}
\end{equation}
where $\b{S}_j$ are basal (b) spins and $\bs{\tau}_j$ are apical (a) spins.
For antiferromagnetic $J_1, J_2$, there is a well-established three-phase structure as a function of the ratio $J_2/J_1$~\cite{BlundellNunezRegueiro03,Jiang2015,RauschKarrasch2025}.
At small $J_2/J_1$ it is in a gapless phase adiabatically connected to the spin-$1/2$ Heisenberg chain.
At intermediate coupling it enters a gapped dimerized phase, including the exactly dimerized valence-bond-solid point at $J_2=J_1$.
For larger $J_2/J_1$, the dimer gap closes into a gapless non-collinear phase, with recent tensor-network estimates placing the transition near $J_2/J_1\simeq1.5$~\cite{BlundellNunezRegueiro03,RauschKarrasch2025}.
This last regime is the same small-$J_1/J_2$ regime on which the present paper focuses.
It is characterized by spiral correlations in the apical sector, crossing over deep in the phase to a commensurate $90^\circ$ spiral with strong weight near momentum $k=\pi/2$~\cite{RauschKarrasch2025}.

\begin{figure}[t]
    \centering
    \includegraphics[width=\linewidth]{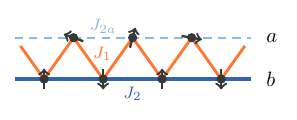}
    \caption{The sawtooth model (solid) and the asymmetric zigzag chain (solid+dashed).}
    \label{fig:model}
\end{figure}

The numerical puzzle stems from a closer look at the properties of this spiral phase.
The calculations of Ref.~\cite{Hu2025} found several signatures that are difficult to organize within an ordinary $z=1$ conformal field theory: the entanglement scaling gives an unstable and anomalously large effective central charge, $c\simeq4.4$; finite-size gaps from exact diagonalization are at least an order of magnitude smaller than in the Heisenberg chain and do not show clean $L^{-z}$ scaling; a simple fitting for a dynamical exponent gives strongly boundary-condition-dependent convergence with inconsistent values of $z$. Furthermore, probing the system with charge fluctuations reveals a nearly flat quasiparticle band with a vanishing charge gap but a finite spatial correlation length.
The spiral phase also shows no detectable dimerization or translation-symmetry breaking \cite{Hu2025}.
Together with the reflection-protected Lieb--Schultz--Mattis (LSM) obstruction discussed there, this absence of detectable symmetry breaking supports a genuine gapless infrared theory. Collectively, these signatures point toward a non-CFT state resembling local quantum criticality \cite{localQCP00, localQCP01, localQCP03}, though its precise nature remains elusive from numerics alone.

In this letter we construct a continuum route into this exotic phenomenology in the weak-interchain-coupling regime.
To understand the difficulties with the sawtooth geometry, we first borrow the standard bosonization approach based on the Emery--Kivelson/Toulouse solution of the Kondo lattice \cite{EK92, ZKE96}, which fails to reach the spiral phase of interest. Directly bosonizing the slow apical sector leads to a multiscale obstruction from the microscopic sawtooth hierarchy, which we avoid by embedding the problem in an auxiliary asymmetric zigzag ladder with two SU(2)$_1$ sectors and strongly unequal velocities.
The resulting two-velocity field theory differs qualitatively from the equal-velocity White--Affleck (WA) problem \cite{WhiteAffleck96} because an unusual marginal twist operator feeds back into the kinetic terms and selectively collapses the slow spin velocity at one-loop level.

The resulting theory reveals an interesting route to the local-critical phenomenology seen in the numerical puzzle: the generated back-scattering channel in the apical chain becomes the leading dimensionless interaction because the slow velocity is driven toward zero.
Meanwhile the dimensionful apical coupling does not show a separated Kosterlitz--Thouless (KT) runaway, and the interchain coupling and basal back-scattering term remain comparatively small.
This naturally accounts for the nearly flat quasiparticle bands, slow dynamical scaling in the whole system, and numerical signatures of usual Heisenberg-chain behavior in the basal spins.
The same flow equations also identify neighboring gapped regimes with dimerization patterns different from the equal-velocity WA phase.

\paragraph*{A Kondo lattice perspective. ---}
We start the continuous approximation by representing the basal spin field as a sum of uniform current ($\b{J}_b^{\rm tot}=\b{J}_b+\bar{\b{J}}_b$) and staggered ($\b n_b$) part: $\b{S}_j = \b{J}_b^{\rm tot}(R) + (-1)^j \b{n}_b(R)$. The sawtooth geometry couples each apical spin to the sum $\b{S}_j + \b{S}_{j+1}$, leading to a pair-wise cancellation of the staggered parts. In Abelian boson fields, the Hamiltonian may be expressed as
\begin{align}
    H_0 &= \frac{v_b}{2}\int dx\,\[\Pi^2 + (\d_x\varphi)^2\],\nonumber\\[4pt]
    H_\p &= -J_\p\,\sqrt{\frac{2}{\pi}}\sum_j\tau^z_j\,\d_x\varphi(R_j),  \label{eq:H_saw_full}\\[4pt]
    H_\perp &= \frac{J_\perp}{\pi a}\sum_j
    \[\tau^+_j\,e^{-i\sqrt{2\pi}\,\vartheta(R_j)}\,
    \cos\!\(\sqrt{2\pi}\,\varphi(R_j)\) + \text{H.c.}\].\nonumber
\end{align}
Here $v_b\sim J_2$ is the spin velocity for the basal chain, and $J_{\parallel} \sim J_\perp \propto J_1$ are introduced to absorb nonuniversal numerical factors.

The form of Eq.~\eqref{eq:H_saw_full} is basically a one-dimensional (1D) Kondo lattice model \cite{kondo77} without the backscattering terms. One can use the Emery--Kivelson transformation $U = \exp[-i\sqrt{2\pi}\sum_j\tau^z_j\vartheta(R_j)]$ \cite{EK92, ZKE96} to exactly cancel the $e^{-i\sqrt{2\pi}\vartheta}$ in $H_\perp$, and modify the $J_{\parallel} \rightarrow J_{\parallel} - \pi v_b$. Hence if we can set $J_{\parallel} = \pi v_b$, all the apical spin operators that are left is $\tau^x$, which becomes just a conserved quantum number, and the resulting theory is well-understood.

In other words, to address the isotropic ($J_{\parallel} = J_\perp$) and weak-coupling ($J_1\ll J_2$) case, it's essential that we bosonize the apical spins as well, which we can justify by making use of the Ruderman-Kittel-Kasuya-Yosida (RKKY) interaction between the apical spins. It turns out that the RKKY interaction takes an interesting form in this particular case, which we now explain. It is useful to rewrite the bosonized Hamiltonian manifestly in terms of the SU(2)$_1$ current $\b{J}_b^{\rm tot}$:
\begin{align}
    H_{\text{SU(2)}} = \frac{\pi v_b}{2}\int dx\,(\b{J}_b^2+\bar{\b{J}}_b^2) + J_1\sum_j \boldsymbol{\tau}_j\cdot\b{J}_b^{\rm tot}.
\end{align}
At second order in $J_1$, integrating out the basal-chain currents produces an RKKY interaction between the apical spins.
Since the current--current correlator decays as $\<J^a(x)J^b(0)\> \propto \delta^{ab}/x^2$, the RKKY coupling is
\begin{align}
    H_{\text{RKKY}} = c_3\frac{J_1^2}{v_b} \sum_{i\neq j} \frac{\boldsymbol{\tau}_i\cdot\boldsymbol{\tau}_j}{|R_i-R_j|^2},
    \label{eq:Hrkky}
\end{align}
where $c_3$ is a non-universal constant of order unity.
This $1/r^2$ spin--spin interaction is precisely the form of the Haldane--Shastry model~\cite{Haldane88, Shastry88}, an exactly solvable SU(2)$_1$ system with interesting quasiparticles even in 1D.

\paragraph*{Asymmetric zigzag embedding. ---}
There is a multiscale problem, with the literal sawtooth estimate $J_{2a}\sim J_1^2/v_b$, also familiar from the Kondo lattice \cite{PhysRevB.76.115108}. The apical chain has bandwidth and continuum cutoff set by Eq.~\eqref{eq:Hrkky}, while the interchain coupling remains of order $J_1$. Thus the coupling that should be treated as a perturbation is already larger than the cutoff of the two-chain Luttinger liquid. A weak-coupling renormalization group (RG) about two decoupled SU(2)$_1$ chains therefore cannot be initialized directly at the microscopic sawtooth hierarchy. To proceed, we assume that the low-energy physics of the sawtooth point is stable over a finite region of asymmetric zigzag ladders with $J_1,J_{2a}\ll J_2$, not only on the singular line $J_{2a}\sim J_1^2/v_b\ll J_1$. The controlled RG is performed in the auxiliary regime $J_1\ll J_{2a}\ll J_2$, where the rung coupling is perturbative below the apical cutoff, and is then extrapolated to diagnose the flow toward the strongly asymmetric sawtooth limit. This local-ladder regularization also bypasses the separate issue that the induced RKKY interaction is long ranged.

We are now ready to set up the RG calculation to determine the low-energy physics for the weak interchain coupling case. Formally, we treat both chains on equal footing, with the only difference being the spin velocities $v_s,\ s=a,b$. After the tree-level relevant back-scattering term is cancelled out, we find \textit{two} leading interactions that are marginal:
\begin{equation}
    H_{\rm int}
    = \int dx\,\Big[
      g_1\,\b{J}_a^{\rm tot}\!\cdot\!\b{J}_b^{\rm tot}
      + g_2\,\big(-\b{n}_a\!\cdot\!\d_x\b{n}_b\big)
    \Big].
    \label{eq:H_int_draft}
\end{equation}
We denote these terms as $\mathcal{O}_1$ and $\mathcal{O}_2$ hereafter. Unlike $\mathcal{O}_1$, the conformal spin of twist interaction $\mathcal{O}_2$ is nonzero since $\d_x \sim \d + \bar\d$. This term is usually ignored in the literature, for example in WA~\cite{WhiteAffleck96} (and in the discussion up to this point in this letter).

To elucidate the physical role of the twist operator $\mathcal{O}_2$, we examine its $z$ component in Abelian bosonization. Using $n^z_s = \cos(\sqrt{2\pi}\varphi_s)$, straightforward calculation shows that, up to total derivatives and boundary terms, $n^z_a\d_x n^z_b$ takes the form
\begin{align}
    \frac{\sqrt\pi}{2}\!\,\Big[(\d_x\varphi_+)\sin(\sqrt{4\pi}\varphi_-) + (\d_x\varphi_-)\sin(\sqrt{4\pi}\varphi_+)\Big],
    \label{eq:O2_zz_pm}
\end{align}
where $\varphi_a\pm\varphi_b = \sqrt{2}\,\varphi_\pm$. We make the observation that this current-vertex product structure may feed kinetic terms rather than simply pinning a field, which is indeed confirmed by the following RG calculation \footnote{Similar operator structures have been discussed in Ref.~\cite{ChenBuettnerVoit01}, but there they work with a different $J_1/J_2$ ratio and a different bosonization starting point. The original operator $(\d_x\Phi)\cos2\Phi$ is later found to be a total derivative and hence not a valid bulk mechanism for new physics \cite{CBV_comment}.}. 

\paragraph*{Two-velocity renormalization group. ---}
The problem we formulate now is the tensor product of two decoupled SU(2)$_1$ Wess--Zumino--Witten (WZW) theories at the self-dual radius $R = 1/\sqrt{2\pi}$, at velocities $v_b$ (basal) and $v_a$ (apical), perturbed by the interaction Eq.~\eqref{eq:H_int_draft}
\footnote{The reason we don't continue with the independent-chain Abelian formulation usually used in ladder analyses such as Ref.~\cite{Ronetti2022} is to avoid the Jordan--Wigner string complication in that approach after each chain is first fermionized separately, at which point extra Klein factors are needed to enforce the interchain commutation relations.}.
The standard operator product expansion (OPE)-based RG (see for example Ref.~\cite{Senechal99}) assumes a single light cone: the OPE kernel $1/|z|^2$, where $z = v_s\tau - ix$, is rotationally symmetric in the Lorentzian coordinates $(\tau, x/v)$, so any radial Wilsonian shell yields the universal $d\ell$ that drives KT.
With two velocities $v_b \ne v_a$ no single radial coordinate makes {all} the OPE kernels rotationally symmetric simultaneously: the cross-species kernels $1/(z_b z_a)$, $1/(z_b\bar z_a)$ live on a hybrid geometry while the stress-tensor kernels $1/z_b^2, 1/z_a^2$ each pick out their own light cone.

To resolve this problem, we choose the basal light-cone as reference and define $v_b\tau = r_b\cos\theta$, $x = r_b\sin\theta$  and integrate over a thin annulus $r_b \in [a, ae^{d\ell}]$ via the element $d\tau\,dx = r_b\,dr_b\,d\theta/v_b$. In this notation, we write the basal coordinates $z_b = r_b e^{-i\theta}$, and the apical coordinates $z_a = r_b\,(\alpha\cos\theta - i\sin\theta)$ with additional parameter $ \alpha \equiv v_a/v_b$ so the usual shell integrals can be performed.

As we now show, this simple scheme reproduces the equal-velocity physics of WA \cite{WhiteAffleck96}, by considering the $\mathcal{O}_1\mathcal{O}_1$ OPE. Using standard SU(2)$_1$ chiral current $\b{J}_s$ OPE with $z_{12} = z_1 - z_2$:
\begin{equation}
    J^\alpha_s(z_1)J^\beta_s(z_2) = \frac{\delta^{\alpha\beta}/2}{z_{12}^2} + \frac{i\eps^{\alpha\beta\gamma}J^\gamma_s(2)}{z_{12}} + :\!J^\alpha_s J^\beta_s\!:\!(2) + O(z_{12}),
    \label{eq:JJ_OPE}
\end{equation}
and that the cross-species commutativity to factorize the product,
\begin{equation}
    \mathcal{O}_1(1)\mathcal{O}_1(2) = \sum_{\alpha\beta}\[J^{\text{tot},\alpha}_b(1)J^{\text{tot},\beta}_b(2)\]\[J^{\text{tot},\alpha}_a(1)J^{\text{tot},\beta}_a(2)\].
    \label{eq:O1O1_product}
\end{equation}
Each bracket contains the four chirality combinations $JJ, \bar J\bar J, J\bar J, \bar JJ$; the last two are regular.
Summing the singular pieces and grouping by operator type,
\begin{equation}
\begin{aligned}
    \mathcal{O}_1(1)\mathcal{O}_1(2)\big|_\text{sing} =\ & \frac{3}{4}\!\(\frac{1}{z_b^2} + \frac{1}{\bar z_b^2}\)\!\(\frac{1}{z_a^2} + \frac{1}{\bar z_a^2}\)\mathbb{1} \\
    & - 2\(\frac{\b{J}_b}{z_b} + \frac{\bar{\b{J}}_b}{\bar z_b}\)\!\cdot\!\(\frac{\b{J}_a}{z_a} + \frac{\bar{\b{J}}_a}{\bar z_a}\) \\
    & + \frac{3}{2}\,T_a^\text{tot}\!\(\frac{1}{z_b^2} + \frac{1}{\bar z_b^2}\) + \frac{3}{2}\,T_b^\text{tot}\!\(\frac{1}{z_a^2} + \frac{1}{\bar z_a^2}\),
\end{aligned}
    \label{eq:O1O1_OPE_full}
\end{equation}
where we define the total stress tensor $T_s^\text{tot} \equiv T_s + \bar T_s$. 

The first line of Eq.~\eqref{eq:O1O1_OPE_full} is only a vacuum-energy renormalization and we will not track it further. 
The second line is the 1-loop generation of the current interaction $\mathcal{O}_1$, but with a chiral splitting from the pole structure: the same-chirality inter-species pieces $\b{J}_b\!\cdot\!\b{J}_a, \bar{\b{J}}_b\!\cdot\!\bar{\b{J}}_a$ on $1/(z_b z_a), 1/(\bar z_b\bar z_a)$, and the opposite-chirality pieces $\b{J}_b\!\cdot\!\bar{\b{J}}_a, \bar{\b{J}}_b\!\cdot\!\b{J}_a$ on $1/(z_b\bar z_a), 1/(\bar z_b z_a)$. 
We therefore group them into two parity-even combinations:
\begin{equation}
    \mathcal{O}_1^\parallel \equiv \b{J}_b\!\cdot\!\b{J}_a + \bar{\b{J}}_b\!\cdot\!\bar{\b{J}}_a,
    \qquad
    \mathcal{O}_1^\perp \equiv \b{J}_b\!\cdot\!\bar{\b{J}}_a + \bar{\b{J}}_b\!\cdot\!\b{J}_a.
    \label{eq:O1_split}
\end{equation}
Note that only $\mathcal{O}_1^\perp$ has a non-vanishing shell integral. Similarly, the third line shows that velocities are not renormalized by the current interaction $\mathcal{O}_1$.
By applying the cumulant $\delta S = -\tfrac{1}{2}g_1^2\int d^2z_2\,(\text{shell-integrated OPE})$, we obtain the following $\beta$-functions
\begin{equation}
    {\quad
    \frac{dg_1^\perp}{d\ell}\bigg|_{(1,1)} = \frac{8\pi\,g_1^2}{v_b + v_a},
    \quad
    \beta_{g_1^\parallel}^{(1,1)} = 0,
    \quad
    \beta_{v_b}^{(1,1)} = \beta_{v_a}^{(1,1)} = 0.
    \quad}
    \label{eq:beta_O1O1}
\end{equation}
As WA noted, the same-chirality current products do not renormalize the interaction at lowest order; only opposite-chirality terms do. To see the gap generation (at equal velocity limit), one may interchange $\b{J}_b \leftrightarrow \b{J}_a$ such that the relevant interaction $\mathcal{O}_1^\perp \rightarrow \sum_{s=b,a}\bar{\b{J}}_s\!\cdot\!\b{J}_s$, reproducing two decoupled spin-$1/2$ Heisenberg chains where KT gaps are generated.

We now turn to the current-twist channel, which renormalizes the twist coupling but does not touch the velocities nor generate gap opening terms. Using the current-primary OPE
\begin{equation}
    J^\alpha_s(1)\,n^\beta_s(2)\big|_{\text{vector}}
    = \frac{i\eps^{\alpha\beta\gamma}n^\gamma_s(2)}{z_s}
    + O(z_s^0),
    \label{eq:Jn_vector}
\end{equation}
direct calculation shows the $\mathcal{O}_1\mathcal{O}_2$ channel factorizes by species similar to Eq.~\eqref{eq:O1O1_OPE_full}:
\begin{equation}
\begin{aligned}
    \mathcal{O}_1(1)\,\mathcal{O}_2(2)\big|_\text{sing}
    =\ & -2\!\(\frac{1}{z_a} + \frac{1}{\bar z_a}\)\!\(\frac{1}{z_b} + \frac{1}{\bar z_b}\)\mathcal{O}_2(2) \\
    & + 2i\!\(\frac{1}{z_a} + \frac{1}{\bar z_a}\)\!\(\frac{1}{z_b^2} - \frac{1}{\bar z_b^2}\)\b n_a\!\cdot\!\b n_b(2).
\end{aligned}
    \label{eq:O1O2_OPE_full}
\end{equation}
The pole structure in the first line again projects out the same-chirality current products $\mathcal{O}_1^\parallel$. We also observe the generation of the relevant singlet $\b n_a\!\cdot\!\b n_b$. However, the kernel is odd in parity and hence produces a vanishing shell integral. We are therefore left with the following $\beta$-functions
\begin{equation}
    {\quad
    \beta_{g_2}^{(1,2)}
    = \frac{16\pi}{v_b + v_a}\,g_1^\perp g_2,
    \qquad
    \beta_{v_b}^{(1,2)} = \beta_{v_a}^{(1,2)} = 0.
    \quad}
    \label{eq:beta_O1O2}
\end{equation}
The $\mathcal{O}_1\mathcal{O}_2$ channel therefore does two things and no more.
It renormalizes the marginal twist coupling $g_2$, and it confirms that parity prevents the generation of the relevant singlet $\b n_a\!\cdot\!\b n_b$ at one loop.
The natural source of genuinely new diagonal kinetic terms or dynamically regenerated backscattering remains the $\mathcal{O}_2\mathcal{O}_2$ channel.

Lastly, the twist-twist channel generates the essential new physics of the asymmetric zigzag chain. The asymmetric density $-\b n_a\!\cdot\!\d_x\b n_b$ is not the best representation for $\mathcal{O}_2$, because the lattice interaction $J_1$ is symmetric between the two chains. To explicitly preserve this symmetry, we use the expression $\mathcal{O}_2 \equiv \frac{1}{2}\(\mathcal{O}_{ab}+\mathcal{O}_{ba}\)$, where
\begin{equation}
    \mathcal{O}_{ab}=-\b n_a\!\cdot\!\d_x\b n_b,\qquad
    \mathcal{O}_{ba}=+\b n_b\!\cdot\!\d_x\b n_a .
    \label{eq:O2_antisym_def}
\end{equation}
They are equivalent up to a total derivative term.
This is also seen in the lattice model, since $\mathcal{O}_{ab}$ and $\mathcal{O}_{ba}$ correspond to reorganizing the $J_1$ interaction sum, up to a boundary term in the action.

The $\mathcal{O}_2(1)\mathcal{O}_2(2)$ product therefore has four pieces, the full expression of which is detailed in the End Matter. Schematically, 
\begin{align}
    \mathcal{O}_{2}(1)\mathcal{O}_{2}(2)\big|_\text{sing}
    \sim \ &
    \Delta\lambda_b\b J_b\!\cdot\!\bar{\b J}_b
    + \Delta\lambda_a\b J_a\!\cdot\!\bar{\b J}_a
    \notag\\
    &+\Delta v_b (T_b + \bar T_b) +\Delta v_a (T_a + \bar T_a)
    \notag\\
    &+(\text{vacuum energy}).
    \label{eq:O2O2_abab}
\end{align}
We can simplify the beta-function contributions sufficiently in the strong velocity asymmetry $\alpha=v_a/v_b\ll1$ limit to see the physics:
\begin{align}
    \beta^{(2,2)}_{\lambda_b}
    &=
    \frac{g_2^2}{v_b}\left[
        2\ln\frac{4}{\alpha}-2
        +O\!\left(\alpha^2\ln\frac{1}{\alpha}\right)
    \right], \label{eq:beta_O2O2_lambda_b}\\
    \beta^{(2,2)}_{\lambda_a}
    &=
    2\frac{g_2^2}{v_b}
    +O\!\left(\alpha^2\ln\frac{1}{\alpha}\right),
    \label{eq:beta_O2O2_lambda_a}\\
    \beta_{v_b}^{(2,2)}
    &=
    -\frac{g_2^2}{v_b}
    \left[
        9\ln\frac{4}{\alpha}-15
        +O\!\left(\alpha^2\ln\frac{1}{\alpha}\right)
    \right],
    \label{eq:beta_vb_O2O2}\\
    \beta_{v_a}^{(2,2)}
    &=
    -3\,\frac{g_2^2}{v_b}
    +O\!\left(\alpha^2\ln\frac{1}{\alpha}\right).
    \label{eq:beta_va_O2O2}
\end{align}
The important structural result is that the twist self-OPE regenerates same-chain opposite-chirality current bilinears, with positive beta-function contributions that feeds into the KT gap channel.
Moreover, velocity flows are also generated via the stress-tensor coefficients $\beta_{v_b}^{(2,2)}$, $\beta_{v_a}^{(2,2)}$ at one-loop level.

\paragraph*{Route to local criticality. ---}
The spin velocity ratio $\alpha=v_a/v_b$ is not itself tied to the usual definition of local quantum criticality.
It is nevertheless a useful proxy for a separatrix between the symmetric WA regime and sector-selective flat-band physics. In particular, the symmetric point means $\alpha = 1$ and $\beta_{\ln \alpha} = 0$ exactly but there is no a priori reason for $\beta_{\ln \alpha}$ to have only one zero. We obtain the ratio flow from Eq.~\eqref{eq:beta_vb_O2O2}-\eqref{eq:beta_va_O2O2}
\begin{equation}
    \frac{d\ln\alpha}{d\ell}
    =
    \frac{g_2^2}{v_b^2}
    \left[
        -\frac{3}{\alpha}
        +9\ln\frac{4}{\alpha}
        -15
        +O\!\left(\alpha\ln\frac{1}{\alpha}\right)
    \right].
    \label{eq:dlnalpha_asym_sanity}
\end{equation}
A striking feature of the full 1-loop $\beta_{\ln \alpha}$ is also reproduced by the strong-asymmetry approximation, which is $\beta_{\ln \alpha} < 0$ for any $\alpha$. This means the velocity ratio $\alpha$ always flows to zero regardless of the degree of the chain-asymmetry in the problem.

The analytic discussion above only follows the direction of the velocity-ratio flow, which describes the velocity collapse in the apical chain for any velocity asymmetry.
To determine how this velocity collapse competes with the remaining $\beta$ functions, we integrate the exact angular integrals numerically.
Physically, $g_2/v_b$ is only a perturbative-control diagnostic: $g_2$ is the coupling that sources the velocity flow and the generated $\lambda$ couplings, but it is not by itself a gap-opening operator.
The quantities with direct strong-coupling interpretations are $g_1^\perp/v_b$, which measures the usual WA/KT current channel, and $\lambda_s/v_s$, which measures the generated intrachain backscattering in sector $s$.

\begin{figure}[t]
    \centering
    \includegraphics[width=\linewidth]{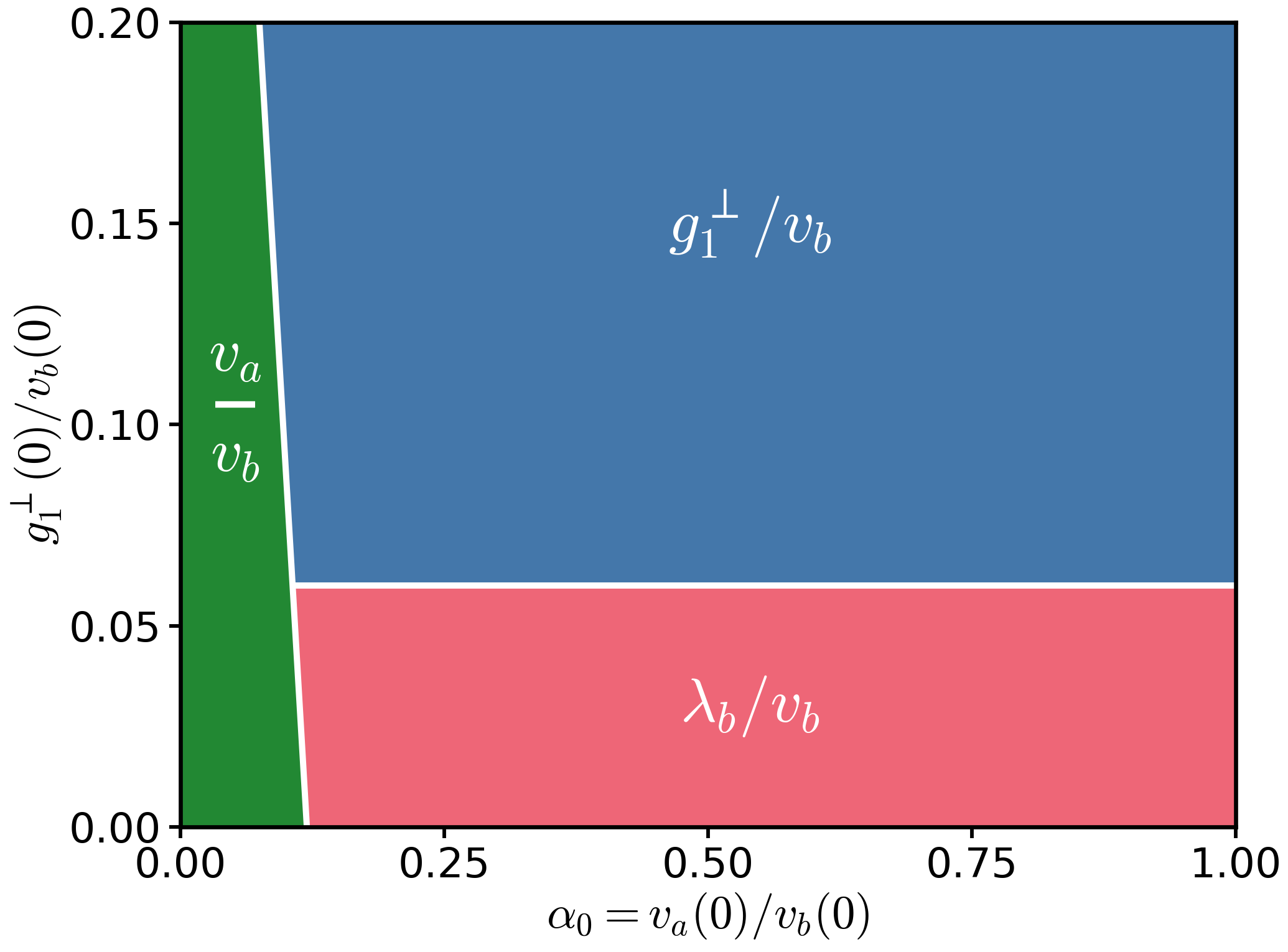}
    \caption{Schematic stopping-event diagram based on the numerical integration of the one-loop RG equations, with $g_1^{\perp}(0)=g_2(0)$. In the green and pink regions, $\lambda_a/v_a$ is the first dimensionless coupling to reach order one.  Green denotes the local-critical candidate regime, where the apical velocity subsequently reaches the velocity cutoff before the basal diagnostic $\lambda_b/v_b$ reaches order one.  Pink denotes the complementary case, where $\lambda_b/v_b$ reaches order one before the velocity cutoff.  Blue denotes WA/KT current-channel domination through $g_1^\perp/v_b$. }
    \label{fig:scan}
\end{figure}

The stopping-event phase diagram in Fig.~\ref{fig:scan} gives the central numerical result, where we stop the flow when either the apical velocity reaches a small cutoff or one of the dimensionless ratios reaches order one.
Even though the exact 1-loop $\beta_{\ln \alpha}$ pushes every asymmetric initial condition away from the symmetric line, the symmetric physics can still win, in the blue region, in the sense that the dominant strong-coupling scale is $g_1^\perp/v_b$ similar to the WA/KT symmetric physics. Due to the generalized LSM theorem \cite{gLSM}, a dimerization must be associated with this gap, whose order parameter is $\<\bs{\tau}_j\!\cdot\!(\b{S}_j-\b{S}_{j+1})\>$ \cite{WhiteAffleck96}.

In the asymmetric region shown in pink/green in Fig.~\ref{fig:scan}, $\lambda_a/v_a$ reaches order one first.
The remaining distinction is what happens closely after if the one-loop equations are continued as a diagnostic beyond $\lambda_a/v_a\sim 1$.
In the pink region, the basal ratio $\lambda_b/v_b$ becomes nonperturbative before the apical velocity reaches the cutoff.
That ordering is closer to an ordinary gap-opening scenario, because the strong coupling occurs in the sector whose velocity has not collapsed. The dimerization order parameter in this case is $\<\b{S}_j\!\cdot\!(\b{S}_{j-1}-\b{S}_{j+1})\>$ in the basal sector, and similar for the apical sector.

Finally, in the green region, the dimensionless ratio $\lambda_a/v_a$ becomes large first because the dimensionful apical scale is tied to a velocity that is itself rapidly collapsing. This is most clearly seen via the strong-asymmetry estimates Eq.~\eqref{eq:beta_O2O2_lambda_a} and \eqref{eq:beta_va_O2O2}.
Approximating $g_2/v_b$ as slowly varying over the initial part of the trajectory and $\lambda_a(0)=0$, then $\lambda_a(\ell) \simeq \frac{2}{3}\left[v_a(0)-v_a(\ell)\right]$. This dictates a profound physical consequence: the spatial scale associated with $\lambda_a$ remains finite while the corresponding energy scale vanishes with $v_a$. We can see this by singling out the Hamiltonian in the apical sector
\begin{align}
    H_a = v_a\[H_{0,a} + \frac{\lambda_a}{v_a}\int dx \b J_a\!\cdot\!\bar{\b J}_a\],
\end{align}
where we explicitly separate the vanishing overall energy scale from the dimensionless coupling driving the spatial correlations.
One can read off the KT correlation length $\xi_a \sim a\exp(c_av_a/\lambda_a) < \infty$ (for $\lambda_a/v_a \lesssim 1$). In contrast, the associated gap $\Delta_a \propto v_a/\xi_a \rightarrow 0$. This is indeed seen in the numerical trajectory in Fig.~\ref{fig:ex_traj}(a). In this sense, our perturbative RG suggests the apical SU(2)$_1$ CFT fails toward a local-critical strong-coupling problem. However, this simple KT picture may not hold as the coupled system enters the strong-coupling regime.

\paragraph*{Discussion. ---}
In summary, we have provided a perturbative RG description for the anomalous gapless spiral phase of the sawtooth spin chain. We make the observation that the essential physics is governed by the extreme velocity hierarchy between the fast basal and (emergent) slow apical spins. By embedding the problem in the asymmetric zigzag chain, this extreme hierarchy activates the marginal twist interaction to drive a sector-selective breakdown of the standard CFT paradigm. Because the apical velocity collapses while its generated backscattering interaction remains finite, the energy scale vanishes independently of the spatial correlation length. While perturbative RG cannot completely resolve the ultimate strong-coupling fixed point, this mechanism provides a possible field-theoretic origin for a local quantum criticality interpretation and for flat-band phenomenology observed numerically, demonstrating why the spiral gapless phase fundamentally escapes the standard equal-velocity WA paradigm.

The present continuum treatment diagnoses the anomalous low-energy scale structure, while the microscopic evolution of the incommensurate spiral wave vector and the effect of charge fluctuations remain open.
A second limitation is that the controlled calculation uses a local asymmetric zigzag regularization, whereas the sawtooth limit naturally induces long-ranged Haldane--Shastry-like apical dynamics; whether this changes the ultimate strong-coupling endpoint is an interesting question for future work.

\begin{acknowledgements}
{\noindent \it Acknowledgements.---} We thank Nick Bultinck and Rui-Zhen Huang for helpful discussions. This research was supported by the European Research Council under the European Union Horizon 2020 Research and Innovation Programme via Grant Agreement No. 101076597-SIESS.
\end{acknowledgements}

\bibliography{ref}

\begin{appendix}

\section*{End Matter}
\section{Representative trajectories}
\begin{figure*}[t]
    \centering
    \includegraphics[width=\linewidth]{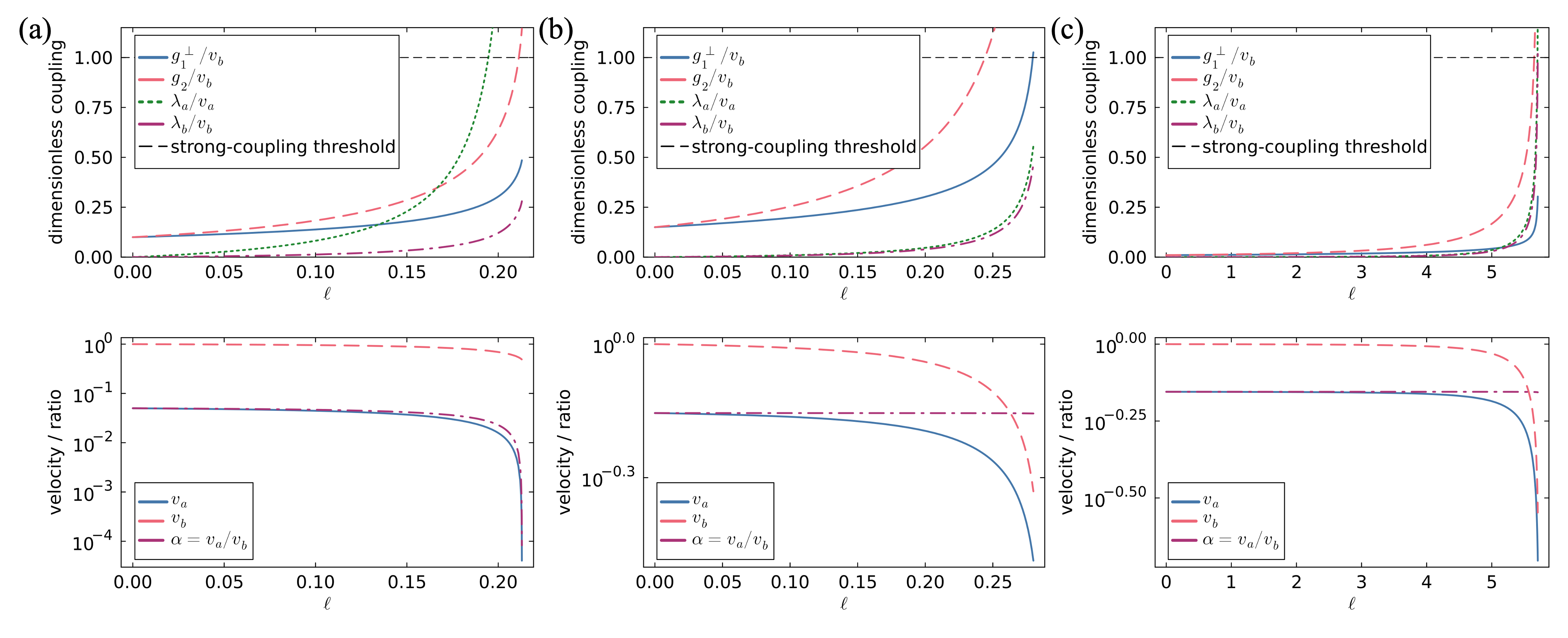}
    \caption{Representative trajectory at (a) $\alpha(0)=0.05$ and $g_1^{\perp}(0)=g_2(0)=0.1$: The apical velocity is driven rapidly downward, while $\lambda_a/v_a$ becomes the leading dimensionless coupling before the basal sector reaches its own strong-coupling scale; (b) $\alpha(0)=0.7$ and $g_1^{\perp}(0)=g_2(0)=0.15$: the WA phase with $g_1^{\perp}/v_b$ the leading dimensionless coupling; (c) $\alpha(0)=0.7$ and $g_1^{\perp}(0)=g_2(0)=0.01$: a similarly gapped phase with both $\lambda_s/v_s$ the leading dimensionless couplings.}
    \label{fig:ex_traj}
\end{figure*}

We show the representative trajectory of each phase in Fig.~\ref{fig:ex_traj}.
The apical velocity decreases and the velocity ratio $\alpha=v_a/v_b$ is rapidly driven toward collapse, while $\lambda_a/v_a$ becomes the dominant dimensionless coupling.
At the same RG time, the basal as well as the WA/KT sector diagnostics remain comparatively small [panel (a)].
This is precisely the sector-selective structure absent with similar velocities [panel (b) and (c)]: the slow apical sector loses perturbative control first, rather than the two chains entering the symmetric WA/KT strong-coupling channel together.

\begin{widetext}
\section{The $\mathcal{O}_2\mathcal{O}_2$ OPE}
\label{sec:O2O2_OPE}
In the $\mathcal{O}_2\mathcal{O}_2$ channel, only the SU(2)-singlet tensor structure is relevant. The coefficients can be determined by matching Abelian bosonization \cite{Senechal99} expressions:
\begin{equation}
    n_s^\alpha(1)n_s^\beta(2)\big|_\text{singlet}
    =
    \delta^{\alpha\beta}
    \Big[
        \frac{1}{|z_s|}
        + \frac{1}{2}\frac{z_s^2}{|z_s|}\,T_s(2)
        + \frac{1}{2}\frac{\bar z_s^2}{|z_s|}\,\bar T_s(2)
        + \frac{1}{3}|z_s|\,\b J_s\!\cdot\!\bar{\b J}_s(2)
    \Big],
    \label{eq:nn_component_new}
\end{equation}
with $|z_s| \equiv \sqrt{v_s^2\tau^2 + x^2}$.
Using the antisymmetrized $\mathcal{O}_2$ with Eq.~\eqref{eq:O2_antisym_def}, the OPE product has four pieces,
\begin{equation}
    \mathcal{O}_2(1)\mathcal{O}_2(2)
    =
    \frac{1}{4}\[
        \mathcal{O}_{ab}(1)\mathcal{O}_{ab}(2)
        +\mathcal{O}_{ba}(1)\mathcal{O}_{ba}(2)
        +\mathcal{O}_{ab}(1)\mathcal{O}_{ba}(2)
        +\mathcal{O}_{ba}(1)\mathcal{O}_{ab}(2)
    \].
    \label{eq:O2O2_product}
\end{equation}
Let $x=x_1-x_2$, $\tau=\tau_1-\tau_2$.
For the same-orientation products the two derivatives act on the same species, so $\d_{x_1}\d_{x_2}=-\d_x^2$ on that species' OPE:
for example,
\begin{align}
    \mathcal{O}_{ab}(1)\mathcal{O}_{ab}(2)
    =\ &
    \sum_{\alpha\beta}
    \Big[n_a^\alpha(1)n_a^\beta(2)\Big]
    \Big[\d_{x_1}n_b^\alpha(1)\d_{x_2}n_b^\beta(2)\Big]
    \notag\\
    \longrightarrow\ &
    \sum_{\alpha\beta}
    \delta^{\alpha\beta}\delta^{\alpha\beta}
    \Big[
        \frac{1}{|z_a|}
        +\frac{1}{2}\frac{z_a^2}{|z_a|}T_a
        +\frac{1}{2}\frac{\bar z_a^2}{|z_a|}\bar T_a
        +\frac{1}{3}|z_a|\,\b J_a\!\cdot\!\bar{\b J}_a
    \Big]
    \notag\\
    &\hspace{2.8cm}\times
    \[-\d_x^2
    \Big(
        \frac{1}{|z_b|}
        +\frac{1}{2}\frac{z_b^2}{|z_b|}T_b
        +\frac{1}{2}\frac{\bar z_b^2}{|z_b|}\bar T_b
        +\frac{1}{3}|z_b|\,\b J_b\!\cdot\!\bar{\b J}_b
    \Big)\].\label{eq:O2O2_abab_step}
\end{align}
Collecting only the dimension-$2$ operators from all four pieces gives the following coefficients:
\begin{align}
    \b J_b\!\cdot\!\bar{\b J}_b:\quad&
    \frac{1}{4}\frac{1}{|z_a|}\[-\d_x^2|z_b|\]
    +\frac{1}{4}|z_b|\[-\d_x^2\!\(\frac{1}{|z_a|}\)\]
    +\frac{1}{2}\(\d_x\frac{1}{|z_a|}\)\(\d_x|z_b|\),
    \label{eq:O2O2_coeff_lambdab}\\
    \b J_a\!\cdot\!\bar{\b J}_a:\quad&
    \frac{1}{4}\frac{1}{|z_b|}\[-\d_x^2|z_a|\]
    +\frac{1}{4}|z_a|\[-\d_x^2\!\(\frac{1}{|z_b|}\)\]
    +\frac{1}{2}\(\d_x|z_a|\)\(\d_x\frac{1}{|z_b|}\),
    \label{eq:O2O2_coeff_lambdaa}\\
    T_b^\text{tot}:\quad&
    \frac{3}{8}\frac{1}{|z_a|}\Re\!\[-\d_x^2\!\(\frac{z_b^2}{|z_b|}\)\]
    +\frac{3}{8}\Re\!\(\frac{z_b^2}{|z_b|}\)\[-\d_x^2\!\(\frac{1}{|z_a|}\)\]
    +\frac{3}{4}\(\d_x\frac{1}{|z_a|}\)\Re\!\(\d_x\frac{z_b^2}{|z_b|}\),
    \label{eq:O2O2_coeff_Tb}\\
    T_a^\text{tot}:\quad&
    \frac{3}{8}\frac{1}{|z_b|}\Re\!\[-\d_x^2\!\(\frac{z_a^2}{|z_a|}\)\]
    +\frac{3}{8}\Re\!\(\frac{z_a^2}{|z_a|}\)\[-\d_x^2\!\(\frac{1}{|z_b|}\)\]
    +\frac{3}{4}\(\d_x\frac{1}{|z_b|}\)\Re\!\(\d_x\frac{z_a^2}{|z_a|}\).
    \label{eq:O2O2_coeff_Ta}
\end{align}
The imaginary parts of the stress-tensor kernels are odd in $x$ and vanish in a parity-symmetric shell, so only the real parts contribute to $T_s^\text{tot}=T_s+\bar T_s$.

Using the coordinate system developed in the main text,
\begin{equation}
    |z_b| = r_b,\qquad
    |z_a| = r_b\,q_a(\theta),\qquad
    q_a(\theta) \equiv \sqrt{\alpha^2\cos^2\theta + \sin^2\theta},
    \qquad
    \alpha \equiv v_a/v_b,
    \label{eq:qa_theta}
\end{equation}
every relevant shell integral again factorizes as $(d\ell/v_b)\int_0^{2\pi} d\theta\,(\text{angular kernel})$.
Writing $c=\cos\theta$ and $s=\sin\theta$, the four shell coefficients are
\begin{align}
    I_{\lambda_b}
    &\equiv \int_\text{shell} d\tau\,dx\,(\text{coefficient of }\b J_b\!\cdot\!\bar{\b J}_b)
    = \frac{d\ell}{v_b}\int_0^{2\pi} d\theta\,
    \left[
        -\frac{c^2}{4q_a}
        +\frac{\alpha^2c^2-2s^2}{4q_a^5}
        -\frac{s^2}{2q_a^3}
    \right],
    \label{eq:Ilambdab_O2O2}\\
    I_{\lambda_a}
    &\equiv \int_\text{shell} d\tau\,dx\,(\text{coefficient of }\b J_a\!\cdot\!\bar{\b J}_a)
    = \frac{d\ell}{v_b}\int_0^{2\pi} d\theta\,
    \left[
        -\frac{\alpha^2c^2}{4q_a^3}
        +\frac{q_a(c^2-2s^2)}{4}
        -\frac{s^2}{2q_a}
    \right],
    \label{eq:Ilambdaa_O2O2}\\
    I_{T_b}
    &\equiv \int_\text{shell} d\tau\,dx\,(\text{coefficient of }T_b^\text{tot})
    = \frac{d\ell}{v_b}\int_0^{2\pi} d\theta\,
    \left[
        \frac{3}{8}\left\{
            \frac{3c^2(c^2-s^2)}{q_a}
            +\frac{(c^2-s^2)(\alpha^2c^2-2s^2)}{q_a^5}
        \right\}
        +\frac{3}{4}\frac{s^2(3c^2+s^2)}{q_a^3}
    \right],
    \label{eq:ITb_O2O2}\\
    I_{T_a}
    &\equiv \int_\text{shell} d\tau\,dx\,(\text{coefficient of }T_a^\text{tot})
    = \frac{d\ell}{v_b}\int_0^{2\pi} d\theta\,
    \left[
        \frac{3}{8}\left\{
            \frac{3\alpha^2c^2(\alpha^2c^2-s^2)}{q_a^5}
            +\frac{(\alpha^2c^2-s^2)(c^2-2s^2)}{q_a}
        \right\}
        +\frac{3}{4}\frac{s^2(3\alpha^2c^2+s^2)}{q_a^3}
    \right].
    \label{eq:ITa_O2O2}
\end{align}

\end{widetext}

\end{appendix}

\end{document}